# Maximizing the Lifetime of Multi-chain PEGASIS using Sink Mobility

Mohsin Raza Jafri[1], Nadeem Javaid[1], Akmal Javaid[2], Zahoor Ali Khan[3]

[1]*CAST, COMSATS Institute of Information Technology, Islamabad, Pakistan.*

[2]*Mathematics Dept., COMSATS Institute of Information Technology, Wah Cant., Pakistan.*

[3]*Faculty of Engineering, Dalhousie University, Halifax, Canada.*

Email: nadeemjavaid@comsats.edu.pk, Phone: +92519049323

*Abstract*— **In this paper, we propose the mobility of a sink in improved energy efficient PEGASIS-based protocol (IEEPB) to advance the network lifetime of Wireless Sensor Networks (WSNs). The multi-head chain, multi-chain concept and the sink mobility affects largely in enhancing the network lifetime of wireless sensors. Thus, we recommend Mobile sink improved energy-efficient PEGASIS-based routing protocol (MIEEPB); a multi-chain model having a sink mobility, to achieve proficient energy utilization of wireless sensors. As the motorized movement of mobile sink is steered by petrol or current, there is a need to confine this movement within boundaries and the trajectory of mobile sink should be fixed. In our technique, the mobile sink moves along its trajectory and stays for a sojourn time at sojourn location to guarantee complete data collection. We develop an algorithm for trajectory of mobile sink. We ultimately perform wide-ranging experiments to assess the performance of the proposed method. The results reveal that the proposed way out is nearly optimal and also better than IEEPB in terms of network lifetime.**

*Keywords*— **Sojourn tour, Sojourn time, IEEPB, Power-Efficient Gathering in Sensor Information Systems**

# I. INTRODUCTION

WSN are used to sense diverse attributes of environment by using wireless sensors. Wireless Routing Protocols strive to ensure efficient data transmission in WSN. As the proficient energy utilization is very important acquisition in WSN, it is possible by using sink mobility in WSNs. Routing protocols, that attempt to ensure proficient energy consumption in WSN (e.g. LEACH [1]), mostly spotlight on the sensors, while a current development specifies a focus shift to the activities of base station (MLMD [2], PNLCM [3]), which can be utilized to advance the network lifetime. As far as, the routing protocols LEACH, IEEPB [4], DEEC [5] and SEP [6] are not sufficient to overcome the energy dissipation of sensors competently so there is a need to utilize sink mobility in WSN to maximize network lifetime. Fixed path mobility is really efficient for its application in enhancing the network throughput. On one hand, a mobile sink can minimize the energy utilization of sensor nodes by collecting information at their place, whereas on the other end it lessens the delay in data delivery for all the nodes in chain. It causes the global load balancing in the entire network of wireless sensors. Wireless routing protocols route the data between the wireless sensors in an optimal way. In mission critical appliances, delay in data delivery is not tolerable such as in forest fire monitoring and nuclear plant monitoring. ERR [7] proposes a design of routing algorithm for delay sensitive relevancies. CBRP [8] portrays importance of cluster-based protocols which always compete with the chain-based protocols performance in case of energy efficiency. So, the amendments in chain-based protocols are necessary to exploit their advantages. Major routing protocol, Power-Efficient Gathering in Sensor Information Systems (PEGASIS) [9] presents the notion of chain formation among sensors and then conveys the data to base station. The leader node collects data of all other chain nodes and sends it to the sink. It is a challenge to achieve well-organized energy utilization of the leader node of chain in PEGASIS and its improved versions but it can be subjugated by sink mobility and multi-head chains. The fixed path mobility minimizes the energy consumption of the sink, but still the main intention of sink mobility remains as the enrichment of network lifetime. It is advantageous in IEEPB as it also saves energy of chain leaders of the multi-head chain. The IEEPB is specially intended for the delay-tolerant applications but the proposed method is also useful for the delay-intolerant applications. It causes tolerable delay in data delivery due to small chains, consisting of fewer numbers of sensors. The multi-chain concept at one hand, decreases the network overhead due to fewer numbers of nodes in chains, whereas on the other hand shrinks the distance between the connected nodes due to their uniform random distribution. Sink mobility alleviates the load on the nodes of chain closer to sink by using the idea of secondary chain heads. The conception of multi-head in the chains diminishes the delay in data delivery and the load on the single chain leader as in IEEPB. The remainder of paper is organized as follows. In Section II, we present related work to our proposed scheme while Section III depicts inspiration for our novelty. Section IV describes the network operations of proposed MIEEPB which includes multi-chain construction, data transmission and sink mobility involving advised algorithm. Section V investigates the simulation outcomes of comparison between our technique and aforementioned proposals. The last section concludes the paper along with expressing the future work.

## II. RELATED WORK

The dilemma of data aggregation and movable sink is extensively scrutinized in the literature [10]. An eminent procedure for sink mobility is Clustering and Set Packing Technique [10]. It deals with the novel scheme, derived from the set packing algorithm and traveling salesman problem. According to the scheme, the sink divides the area into clusters on the basis of Set Packing Algorithm and determines the route between cluster heads via traveling salesman algorithm. EART [11] recommends multi-path routing idea, which assures improved quality of service (QoS) by examining parameters such as reliability, timeliness and energy in WSNs. DAMLR [12] is a constructive framework that uses well-studied linear programming and Lagrangian method to design maximum lifetime algorithm. It is able to achieve distributed algorithm based on the sub-gradient method to determine sink sojourn times and routing flow vectors to sink sojourn locations. The major limitation in the above approaches is that, they do not specify the design of sink mobility for routing protocols such as LEACH, SEP and PEGASIS.

The afore-mentioned proposals discusses the movement of sink for benefit of micro-sensor nodes (MSN) whereas, IEEPB modifies the process of chain formation and leader selection in EEPB. It removes the long link (LL) problem using threshold computations. Along with sink mobility, Data compression and sampling techniques are significant solution to lengthen the life span of WSNs. With the deployment of large sensor networks, the crisis of data sampling and compilation is becoming serious. In-network compression techniques are suitable solutions to tackle this problem. DCS [13] suggest an innovative algorithm for in-network compression, based on distributed source coding (DSC) and compressive sampling (CS) aiming at longer network lifetime. Every node separately takes a decision regarding the compression and data promotion plan to reduce the quantity of packets to convey. In this procedure, the adaptive algorithm based on discrete cosine transform (DCT) and Pack-and-forward (PF) scheme is highly constructive in acquiring the optimal compressibility rate. In wireless multimedia sensor networks (WMSN) domain, QRP [14] uses genetic algorithm (QGA) and queuing theory to assess the path's quality of service. It allocates a weight to quality of services on the basis of hindrance, consumed energy and consistency and thus opts for the top path.

## III. MOTIVATION

In recent years, researchers propose numerous chain based protocols such as PEGASIS, EEPB [15] and IEEPB et al. Among them, IEEPB reduces the construction of long link (LL) in chain by using threshold. Furthermore, it employs the residual energy and the calculated weight of sensor nodes to compute the chain head of the chain. Latest proposals show that sink mobility is highly advantageous in enhancing the network lifetime. MLMD and PNLCM are two prominent research works that verify the effectiveness of sink mobility in MSNs. In PEGASIS, network lifetime is considered as a sole metric for competence but minimization of data delivery delay is also an important acquirement. JSMR [16] not only advises an outline to augment the life span of a WSN, but also attempt to prevail over delayed data delivery to base station by taking advantage of sink mobility. In the domain of multi-path routing, recent researches show that there is a space for enhancements and evolution. REMP [17] proposes a new technique, guaranteeing high quality of service utilizing multi-path routing. In our technique, we also exercise multi-path routing by concept of multi-headed chains. IEEPB removes several deficiencies in EEPB; however, it has still some imperfections as follows:

- When IEEPB builds a chain, there is a major load on the single chain leader due to distance between sink and itself, which causes much energy consumption.
- A large delay in data delivery exists due to long single chain so it is not well suited for delay-sensitive applications.
- The sparse nodes of the network in instability period are badly affected due to long mutual distances.

In this paper, we investigate the problems of long chain for efficient energy consumption and delay in data delivery considering sink mobility. We spotlight on the notion of fixed path mobility and confine the sink to limited potential locations. We also diminish the load on the single chain leader by launching the notion of multi-head chains. On the basis of afore-mentioned investigation, this paper presents mobile sink improved energy-efficient PEGASIS-based routing protocol (MIEEPB). MIEEPB introduces the sink mobility in the multi-chain model, therefore achieving smaller chains and decreasing load on the leader nodes.

## IV. NETWORK OPERATIONS OF THE PROPOSED MIEEP

The main steps in operations of MIEEPB are following.

### A. Network model

We consider a 100m x 100m area for WSN. In our scenario, 100 nodes are deployed in which 25, 25 nodes are further divided randomly in equally spaced area using uniform random distribution. We assume sink mobility as the sink moves through the centre of equally spaced regions and complete its full trajectory in 1 entire round. Sink stays at sojourn location for specific time duration known as sojourn time. Like IEEPB, we ignore the effect of signal interference in wireless channel. We employ first order radio model to calculate energy consumption in data transmission by sensors.

$$E_{tx}(k,d) = E_{tx}\text{-}elec(k) + E_{tx}\text{-}amp(k,d) \tag{1}$$

$$E_{rx}(k) = E_{rx}\text{-}elec(k) \tag{2}$$

$$E_{DA}(k) = E_{DA}\text{-}elec(k) \tag{3}$$

According to first order radio model, $E_{elec}$ = 50 nJ/bit is consumed by the radio to run the transmitter or receiver circuitry and $E_{amp}$ = 100 pJ/bit/m2 is required for transmitter amplifier, where $k$ are number of bits and $d$ is distance. Transmitter circuitry also consumes $E_{DA}$ = 50 nJ/bit to aggregate the data received by the child nodes. The sink divides the area into 4 equal regions in which nodes are deployed by uniform random distribution. The sink moves in these regions 1 by 1, with the particular speed and completes its course in a round. Each sensor compresses the received bits by a data aggregation (*DA*) factor of 0.6 using distributed compressive sampling.

*Figure1*

B. Multi-chain construction

The procedure of chain building is same as of PEGASIS. In MIEEPB, there are 4 chains in our proposal so chain formation occurs in following way.

- Sink sends hello packet to all the nodes to get information of all the nodes.
- Sink finds the farthest node by comparing the distances of all the nodes from itself in first region.
- The chain formation starts from the farthest node *i* also known as end nodes. The end node finds the nearest node from itself.
- Therefore, each node finds the distance between itself and the nearest node not connected in chain, and then connects with it following the same approach.
- In the chain, each node *i* receiving data from the node j, acts as a parent to node *j*, whereas node *j* acts as a child to node *i*.

The same process of chain formation repeats in all 4 regions and thus, 4 chains are created.

C. Chain leaders' selection

In this section, chain chooses the first chain leader on the basis of weight *Q* assigned to each node. Each node computes its weight *Q* by dividing its residual energy with its distance from the base station. The network compares the weights of all the nodes in chain. The network computes the node having highest weight and judges it as primary chain leader of the chain. After the chain formation, each node $i$ computes its distance $d_p$ with the parent node and then, compares it with the distance $d_{bs}$ to the sink. If the later distance is less ($d_p \geq d_{bs}$), the node $i$ acts as a secondary chain head and sends the collected data to the sink, instead of transferring it to parent node.

$$Qi = Ei/Di \qquad (4)$$

where $E_i$ denotes the residual energy of sensor node $i$ while $D_i$ indicates the distance between sensor node $i$ and sink.

*D. Sink mobility*

We consider that sink has unlimited amount of energy and its mobility is used to maximize the network lifetime. Sink moves in WSN in a fixed trajectory, travels from one region to the other and waits for a sojourn time at sojourn location. Sojourn time is the time interval for which sink stays at specific position and gets data from the chain leaders. Sojourn location is the location where the sink temporarily stays for data collection. In our proposed setting, the sojourn locations are (33m, 25m), (33m, 75m), (66m, 25m) and (66m, 75m).

1) *Proposed algorithm for sink mobility:* We advise a scalable algorithm for the distance constrained mobile sink to deal with the above mentioned problems. It consists of following three stages. The sojourn time profile at each sojourn location is calculated first. Based on the sojourn time profiles, it then starts sojourn tour for the mobile sink by identifying the sojourn locations of (33m, 25m), (33m, 75m), (66m, 25m) and (66m, 75m) subject to the specified constraints. In third step, it calculates the total sojourn time in 1 round. The sink computes the total sojourn time of all 4 sojourn locations in 1 round as follows

$$T_s = \sum_{i=1}^{4}(\tau_i) \qquad (5)$$

where $T_s$ is the total sojourn time of 1 course. The goal (1) is to augment the network duration by enhancing the total sojourn time, whereas the objective (2) is to guarantee consistent data transmission among the sink and chain leaders. Therefore, the problem is to

$$\text{Maximize} \quad \sum_{i=1}^{4}(\tau_i) \qquad (6)$$

subject to:

$$x_{ij} = \begin{cases} D & if\ i = j \\ 0 & otherwise \end{cases} \qquad (7)$$

where $x_{ij}$ is the number of bits transmitted between chain leaders and the sink having potential locations *i* and *j*, $1 \leq i, j \geq 4$. D is the total data transferred between chain leaders and the sink in sojourn time. Potential locations depend on the 4 uniform regions. The chain leaders of the same chain, conveys data to the sink at single sojourn location using TDMA.

*E. Data transmission*

MIEEPB uses the same technique for data transmission as in IEEPB. Chain leader transmits token to the end nodes. As the end node sends its data, it passes the token to the next node. Therefore, following the same process, each node *i* after sending its data transmits token to the next node *i-1* in the chain. Thus, all the nodes in the chain transmit their data to the chain heads and the chain heads sends data to the sink. As the sink moves into the region of first chain, it receives data from chain leader s of the first chain and the same process repeats in all the 4 regions. Therefore, data transmission in MIEEPB is based on the token passing approach in which, token passing starts from the end nodes towards leader nodes of the chains. If there is more than 1 child of any node *i*, TDMA mechanism is used for data transmission. The smaller chains in multi-chain scheme chiefly shrink the delay in data delivery.

*F. Data aggregation using DCS*

Each node *i* receives the data of its child node and compresses it using DCT as proposed in DCS [10]. It combines its data with the received one using compressive sampling. Therefore, the received data is compressed by a data aggregation factor.

## V. SIMULATIONS AND RESULTS

In this section we assess the performance of sink mobility in multi-chain IEEPB using MATLAB. We consider a 100m x 100m area for WSN. Our proposed scenario consists of 100 nodes, in which 25, 25 nodes are further divided arbitrarily in equally spaced 4 regions. Sink mobility is proposed as the sink moves about the centers of equally spaced regions and complete its course in 1 round. The simulation parameters are given in Table 1.

*Table 1*

Now, we compare network lifetime of MIEEPB with IEEPB in homogeneous network. We carry out simulations for 5000 rounds. For this comparison, we perform independent simulations (i-e starting from different random number seeds). Figure 2 represents the number of alive sensor nodes during the network duration. Comparison shows that network lifetime and stability period of MIEEPB is better than that of IEEPB, which are approximately 4300 rounds and 1500 rounds respectively. It is because; in MIEEPB, chain leaders die more slowly due to sink mobility. Unlike IEEPB, MIEEPB has longer instability period which is around 2800 rounds as shown in Figure 2. The time duration between the death instants of the first alive node and last alive node in the network is considered as instability period. In our proposed scenario, instability time is better than of IEEPB, because energy distribution is done efficiently. Resourceful utilization of energy becomes possible due to modification of chain in every round.

*Figure 2*

In MIEEPB, first node dies at about 1500 round which is far better than the stability period of IEEPB. Furthermore, the lifetime of first 10 nodes is much improved than of the afore-mentioned techniques due to reduction of load on the chain heads, thus causing global load balancing in the network. In multi-chain concept, distances between the connected nodes are less than single chain therefore; energy consumed in data transmission is less than the single chain. Residual energy of network in MIEEPB decreases more slowly than in IEEPB. In MIEEPB, network lifetime is 79% better than the earlier technique due to efficient energy utilization. The whole network dies at 2400 rounds in IEEPB while in our proposed method network dies at 4300 rounds, so instability time in MIEEPB is 86 % more than the other protocols. The multi-head chain model removes the long link (LL) problem by sending data directly to the sink in case of remote parent node. It further diminishes the delay in data delivery to base station. Figure 3 shows the assessment of MIEEPB and IEEPB in terms of the dead nodes. In IEEPB, the number of dead nodes undergoes sudden increase due to the long link problem in the chain after about 1600 rounds. At that instant, the distant nodes consume much energy due to long link problem after the dying of first 25 nodes. Because of minor chains in our proposed scheme, there are not much longer distances between the remaining nodes of the multi-chains after the dying of first 25 nodes in all 4 regions. There is a less data delivery delay in leader and the sink is smaller in last 1000 rounds due to the sink mobility.

*Figure 3*

As the network lifetime of MIEEPB is significantly greater than the earlier one, it means that the nodes transmits more packets to BS (i-e the throughput is high). During last 1000 rounds of instability period, nodes density is notably low, a lot of empty spaces (in term of the coverage) are formed, due to which network gets sparse. In spite of large empty spaces, our proposed technique provides better coverage in last 1000 rounds than of IEEPB, because of this BS receives additional packets in our proposed scenario. From Simulation results in Figure 3, we estimate that the instability period of MIEEPB largely increases due to the sink mobility and provides better coverage. Furthermore, it is better for the delay sensitive applications due to smaller chains. In our simulations, all the nodes have equal amount of initial energy of 0.5 joules. The nodes are termed as dead, if they losses all of their energy therefore, they drop transmitting or receiving capabilities. Figure 4 presents the contrast of energy consumption of MIEEPB with other protocols.

*Figure 4*

The energy distribution on both scenarios is almost equal. Simulation result shows that the residual energy of network over rounds decreases gradually in both techniques, but there is much efficient energy consumption in our proposed technique. In IEEPB, cluster head sends the data to sink and the larger amount of energy is utilized due to long distance between chain leader and sink. In MIEEPB, chain leaders consume less energy for this purpose due to sink mobility which causes less distance between the chain leaders and the sink. Figure 5 presents the comparison of normalized average energy consumption of the MIEEPB with other protocols.

*Figure 5*

Simulation results show the normalized average energy consumption of sensor nodes over rounds in MIEEPB is 2 % better than the previous methods. The distance among sparse nodes themselves and the base station is fewer than in IEEPB; this practice saves plenty of energy.

## VI.   CONCLUSION AND FUTURE WORK

In this paper, we recommend a multi-chain model of PEGASIS along with induction of sink mobility to maximize the network lifetime. Our considerations are supportive in diminishing the delay in data delivery and distances between the connected nodes through smaller chains. Sink mobility not only lessens the load on the chain leaders in starting rounds, but also reduces the stress on the sparse nodes at the end of network lifetime. We also propose an algorithm for fixed path sink mobility in our model. Sink mobility has major advantages on static sink in enhancing the network lifetime. As for future directions, we are striving to get much better sink mobility specifically toward chain leaders of chains in WSN.

# Tables in the paper of MIEEPB

Table 1

Parameters used in Simulations

| Parameter | Value |
|---|---|
| Network size | 100m x 100m |
| Node number | 100 |
| BS sojourn locations | (33m,25m),(33m,75m),(66m,25m),(66m,75m) |
| Initial energy of nodes | 0.5J |
| Packet size | 2000 bits |
| BS location in IEEPB | (0m,0m) |

# FIGURES

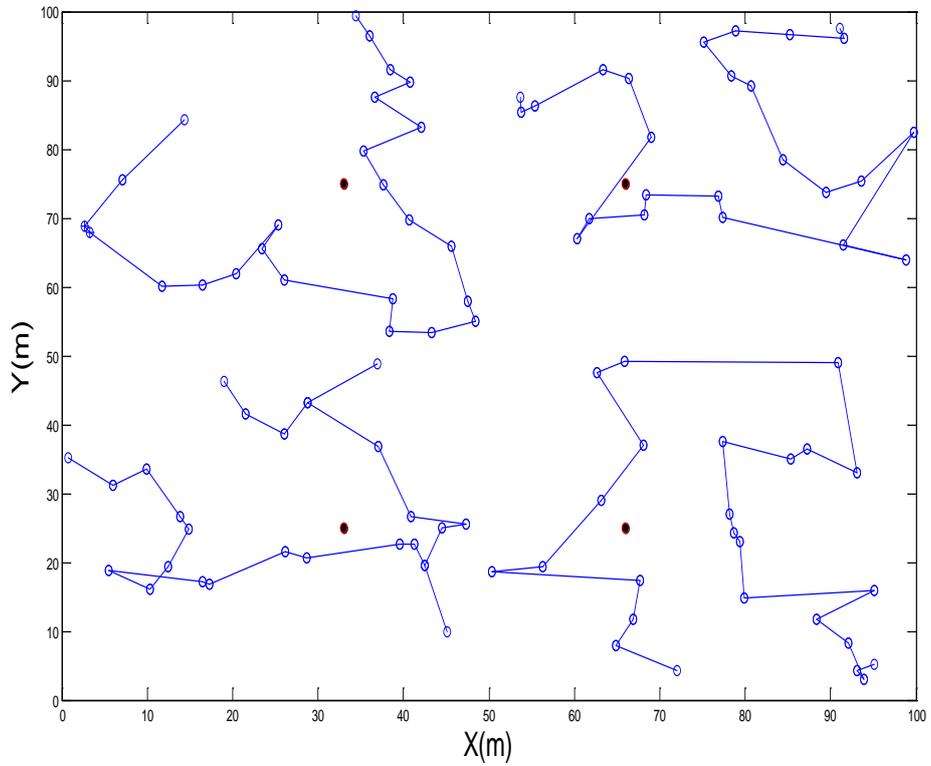

Figure. 1. Chains formation and Sink Sojourn Locations in MIEEPB

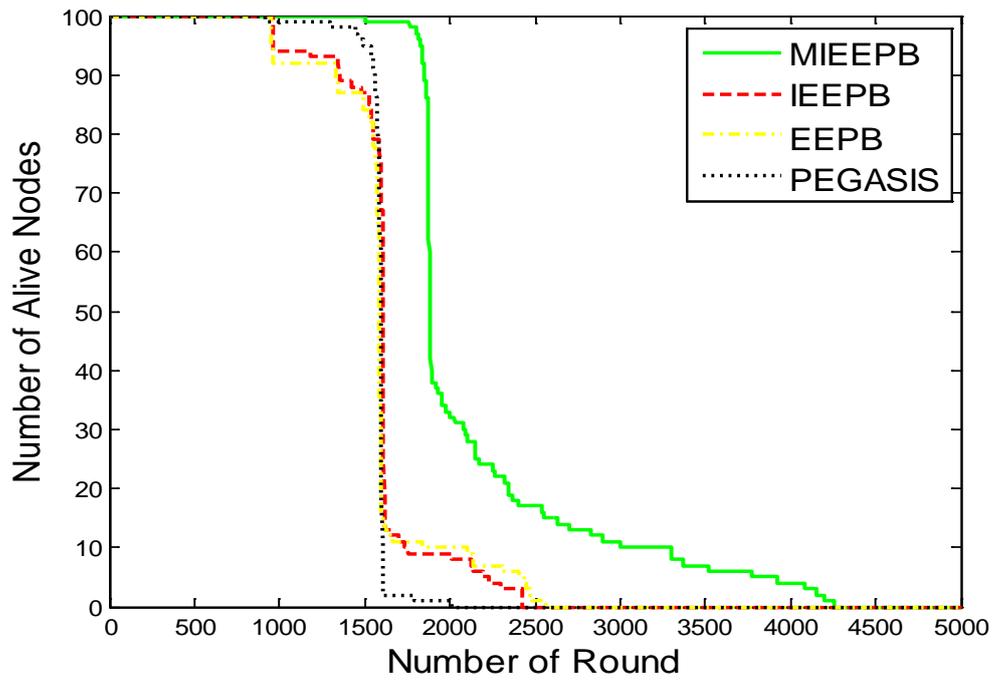

Figure. 2. Network Lifetime Graph

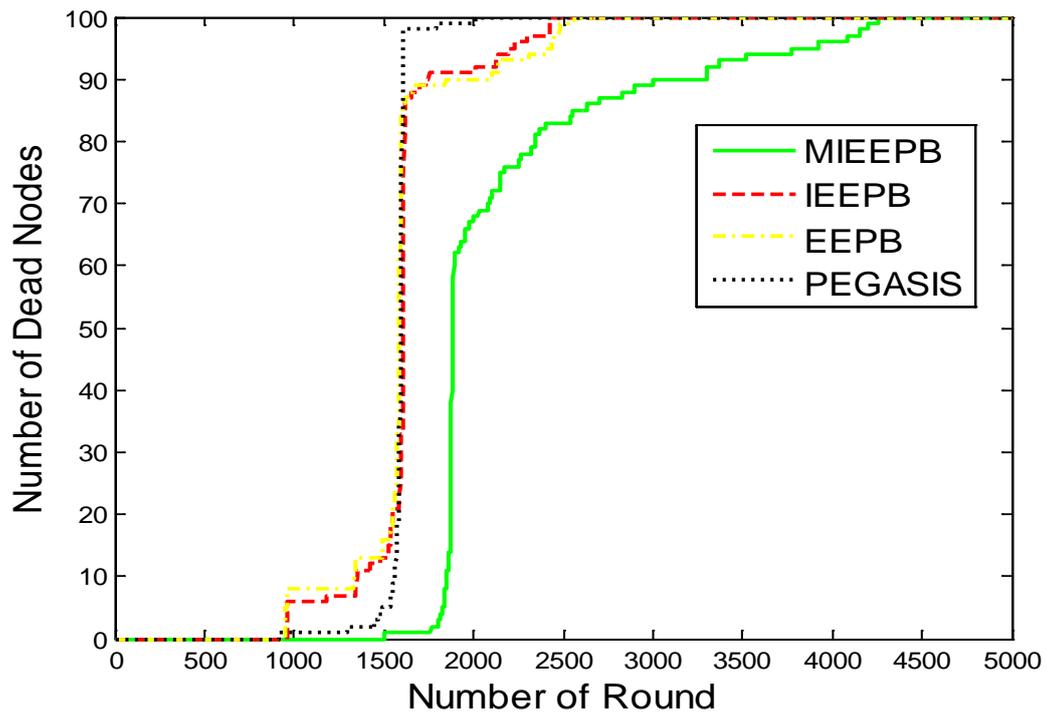

Figure. 3. Comparison of Dead nodes in MIEEPB, IEEPB, EEPB and PEGASIS

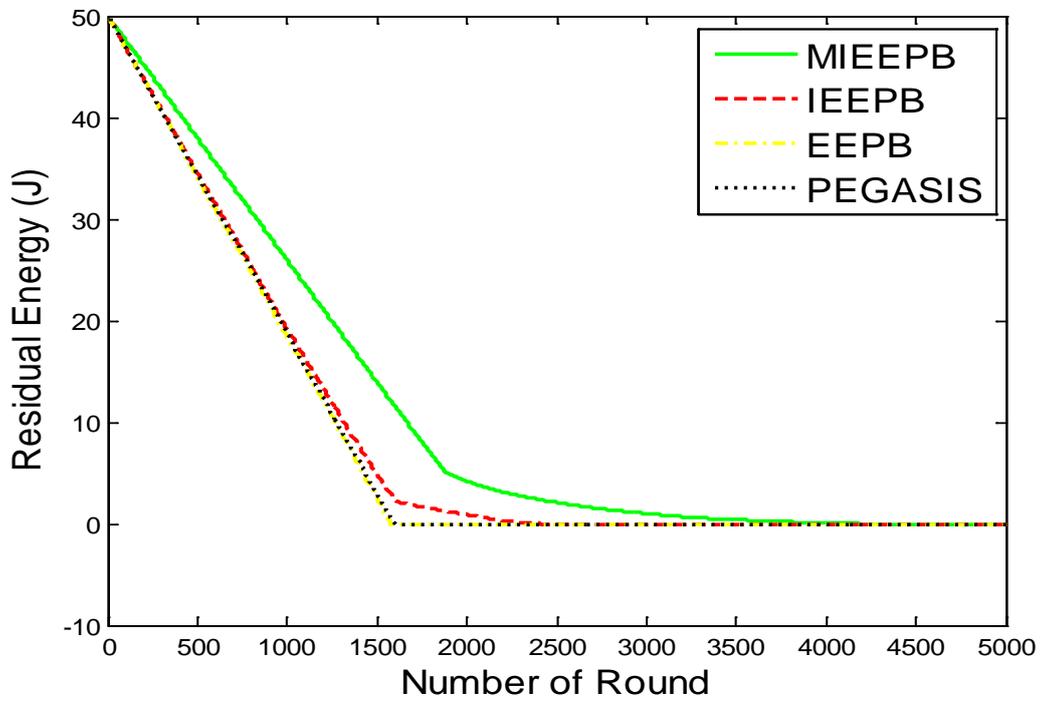

Figure. 4. Comparison of Energy Consumption in MIEEPB, IEEPB, EEPB and PEGASIS

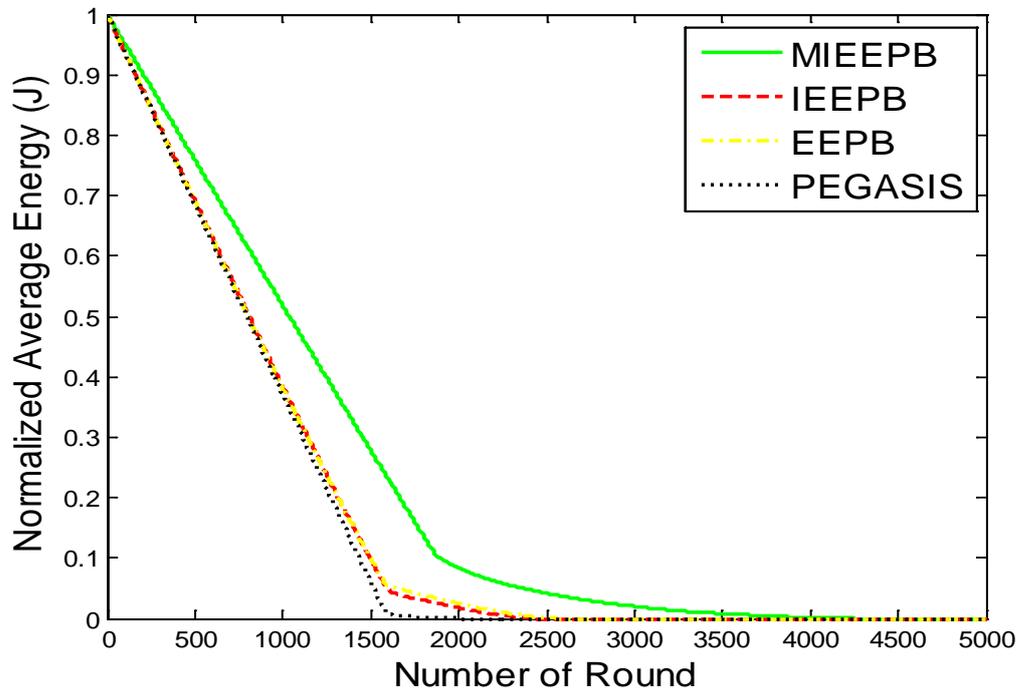

Figure. 5. Comparison of Normalized Average Energy Consumption in MIEEPB, IEEPB, EEPB and PEGASIS